\documentclass[aps,reprint,amsmath,amssymb,groupaddress,superscriptaddress,pra]{revtex4-2}

\usepackage[utf8]{inputenc}
\usepackage{amsfonts}
\usepackage{amssymb}
\usepackage{bm}
\usepackage{graphicx}
\usepackage{soul}
\usepackage{physics}
\usepackage{multirow}
\usepackage{comment}
\usepackage{textcomp}

\usepackage[colorlinks,citecolor=blue,linkcolor=red,anchorcolor=blue,urlcolor=blue]{hyperref}
\usepackage{dcolumn}
\usepackage{physics}
\usepackage{xcolor}
\usepackage{floatrow}
\usepackage{newunicodechar}
\newunicodechar{，}{,}
\usepackage{dcolumn}
\usepackage{color}
\usepackage[T1]{fontenc}
\usepackage{booktabs, array, mathptmx, float, tabularx, lipsum, amsmath, multirow,mathtools}
\usepackage{CJK}
\usepackage{placeins}
\begin{document}

\title{Stability of rotating magnetic levitation}

\author{Mingjun Fan}
\address{School of Physics and Astronomy, Beijing Normal University, Beijing, 100875, China}
\affiliation{Beijing National Day School, Beijing, 100039, China}
\author{Jinyu Chen}
\affiliation{Beijing National Day School, Beijing, 100039, China}
\author{Yongquan Ji}
\affiliation{Beijing National Day School, Beijing, 100039, China}
\author{Long Li}
\affiliation{Beijing National Day School, Beijing, 100039, China}
\author{Chichuan Ma}
\address{School of Physics and Astronomy, Beijing Normal University, Beijing, 100875, China}
\affiliation{Beijing Navigation School，Beijing, 101117, China}
\author{Yu-Han Ma}
\email{yhma@bnu.edu.cn}
\address{School of Physics and Astronomy, Beijing Normal University, Beijing, 100875, China}
\address{Key Laboratory of Multiscale Spin Physics (Ministry of Education), Beijing Normal University, Beijing 100875, China}
\address{Graduate School of China Academy of Engineering Physics, Beijing, 100193, China}

\date{\today}

\begin{abstract}
Dynamical magnetic levitation has attracted broad interest in the realm of physics and engineering. The stability analysis of such system is of great significance for practical applications. In this work, we investigate the stable magnetic levitation of a floater magnet above a rotating magnet and copper board system. The conditions for stable levitation are analyzed through both theoretical modeling and experimental observation. This study focuses on the interplay between magnetic forces, damping effects from the copper board, and rotational dynamics. We derive the equilibrium conditions, perform stability analysis, and present phase diagrams in parametric spaces of rotation speed and damping coefficients. The theoretical predictions show qualitative agreement with experimental results, particularly in demonstrating how damping is essential for stable levitation and how the stability region depends on the geometric and magnetic parameters of the system.
\end{abstract}

\maketitle

\section{Introduction}
Magnetic levitation, a key technology that enables frictionless suspension and precise, contactless manipulation, has found applications in diverse fields, ranging from transportation and materials science to biomedical engineering~\cite{turker2018_magLev1,ge2020_magLev2,Ge2020MagLevApp}. However, Earnshaw’s Theorem prohibits stable configurations of ferromagnets solely by static magnetic force. One well-known way to circumvent this limitation is through special properties of materials, such as diamagnetic ~\cite{simon2001diamag} and superconductive materials~\cite {supreeth2022_superconReview}. However, magnetic levitation can also be realized through classical mechanics only, as Earnshaw's Theorem might not restrain dynamical systems. 

In fact, dynamic magnetic levitation systems have been richly explored over the last century, exemplified by the famous toy called Levitron~\cite {simon1997_levitron, perez2015_drivenLevitron}, the magnetic stir~\cite{baldwin2018_flea}, and Magnetic Paul Trap systems~\cite{perdriat2023_paulTrap}. In these systems, rotation introduces non-inertial effects that confer stability~\cite{Kirillov2021nonConStab}, akin to phenomena observed in celestial mechanics, such as stability near Lagrange points~\cite{farquhar1973_quasi}) and in classical mechanical systems, such as the rotating saddle~\cite{kirillov2016_saddle}.

A recently identified subclass within these dynamical levitation systems, first reported in detail by Ucar in 2021~\cite{ucar2021_polarity}, distinguishes itself from all those well-studied phenomena in that the configuration of the floater magnet is unintuitively orthogonal to the external magnetic field during stable levitation rather than being parallel to it. Such unusual orthogonal configuration, besides gaining interest in the general public, ~\cite{IPT2023,IYPT2024,youtube2021maglev} enables new modes of contactless ferromagnetic particle manipulation besides existing ones ~\cite{ge2020_magLev2,turker2018_magLev1}, particularly in environments where viscous or eddy-current-induced damping is present, as this particular phenomenon requires damping for stable levitation. 

Building on Ucar’s foundational work, Hermansen et al.\cite{hermansen2023_rot} implemented systematic experiments with quantified observations about levitation and compared the results qualitatively with their simulation. More recently, Le Lay et al.\cite{le2024_stable}, by reducing degrees of freedom, offered a clean physical picture about stability in the z direction and derived scaling laws about how the magnetic moment of magnets and rotational speed affect nutation angle and levitation height, which was confirmed by experimental observations.

Besides balance position, the stability of balance position is of great significance. This is because in practical situation, the system is subject to perturbations, under which the floater will oscillate around the balance position if stable, diverge from the balance position if unstable ~\cite{merkin2012_introduction}. The boundary of the stable region in parametric space has remained underexplored in numerous physical and engineering systems, ranging from satellite orbiting~\cite{bardin2008SatOrbital}, plasma confinement~\cite{wesson1978plasmaStab,balbus1992magDisk}, structural vibration control~\cite {meerkov1980VibControl}, robotics manipulation~\cite{murray2017robControl}, and magnetic levitation is of course no exception. However, to the base of our knowledge, previous studies have not rigorously addressed the stability of the floater, which restricts the practical application of such system. They simplified the degrees of freedom and neglected damping effects induced by nearby conductive materials—effects that are experimentally observed to be essential for achieving levitation. This leads to previous studies failing to consider the system as both highly coupled and dissipative, two properties that we believe are significant for stability. To comprehensively analyze the stability of levitation, in this work, we retain all six degrees of freedom of the levitating magnet and explicitly incorporate eddy current damping effects.

The remainder of this paper is structured as follows: In Sec.~\ref{Sec.Modeling}, we analyze the equilibrium configuration of the floater in a co-rotating non-inertial reference frame, wherein the rotor remains stationary, and discuss the spatial symmetries of the system. Section~\ref{Sec.Stability Analysis} is devoted to the stability analysis of the equilibrium configuration. We construct a stability diagram in the parametric space of the damping coefficient and rotor angular velocity. In Sec.~\ref{Sec:expVerification}, we compare theoretical predictions with experimental observations. Finally, Sec.~\ref{Sec.conclusion} provides a summary and discussion of the implications of our findings.

\begin{figure}
    \centering
    \includegraphics[width=0.7\linewidth]{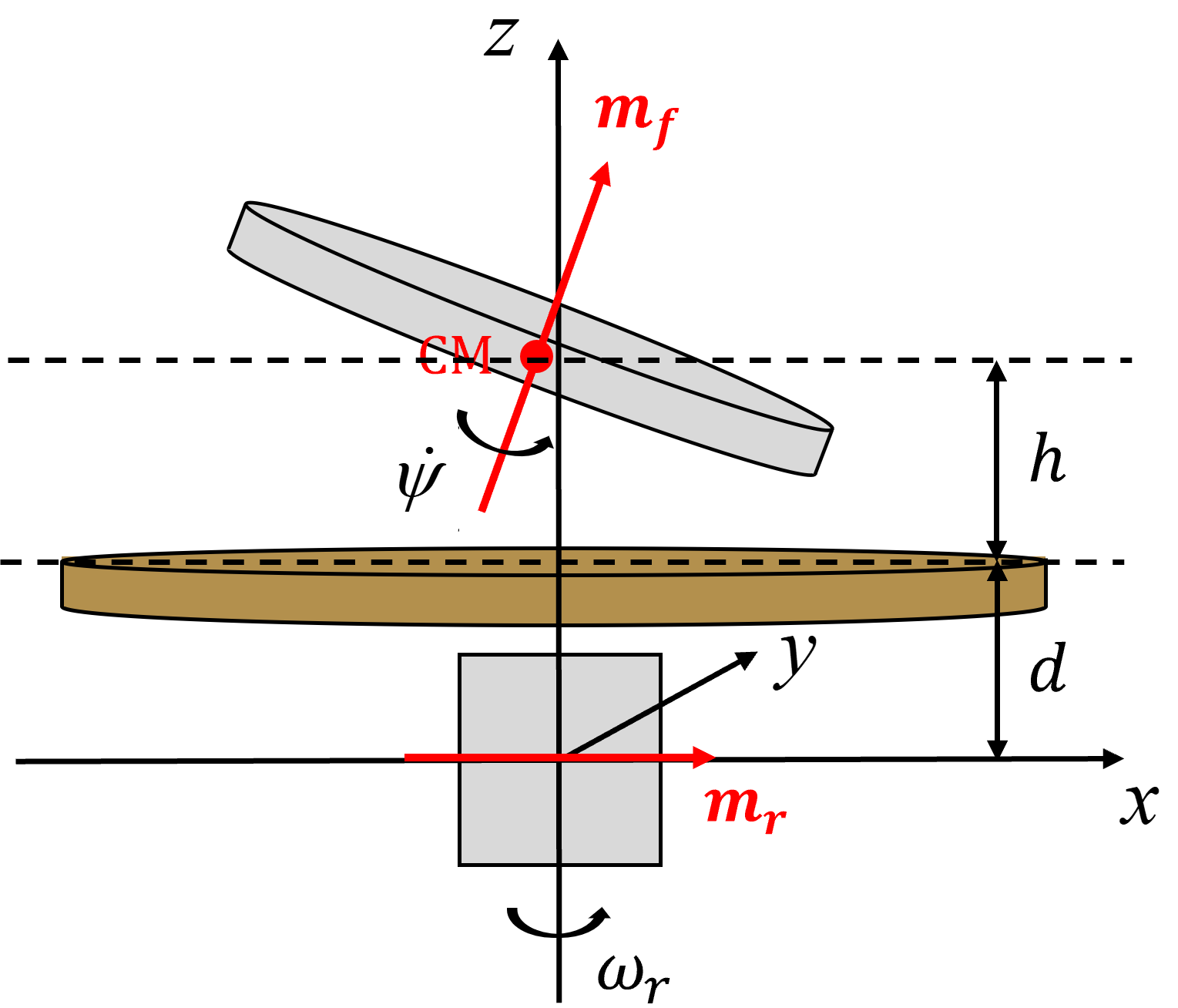}
    \captionsetup{justification=justified,singlelinecheck=false}
    \caption{Sketch of the system}
    \label{fig:coordinateDef}
\end{figure}

\section{Modeling}
\label{Sec.Modeling}
As shown in Fig.~\ref{fig:coordinateDef}， the setup to be explored comprises three primary components: a rotating magnet (rotor)-with magnetic moment $\mathbf{m_f}$ and angular velocity $\omega_r$- positioned at the bottom, a levitating magnet (floater) -with magnetic moment $\mathbf{m_r}$, mass $m$, and moment of inertia about the axis perpendicular to the symmetry axis $I_{12}$ - placed above, and a copper board placed between them. We note that in previous studies ~\cite{hermansen2023_rot,le2024_stable}, the rotor is slightly tilted rather than completely horizontal because their floater is beneath the rotor, and the tilting creates an upward magnetic force for the floater to balance gravity. Here, we set the tilting to be zero because our floater is above the rotor, and such a setting offers a cleaner picture in stability analysis. 

The starting point for our theoretical modeling based on the following two experimental observations: i) At low $\omega_{r}$, levitation does not occur. As $\omega_{r}$ increases, the floater begins to levitate and exhibits precessional motion around the $z$-axis with a small nutation angle that decreases as $\omega_{r}$ increases; ii) The floater processes synchronously with the rotor with zero phase lag. A comprehensive depiction of our experiments is given in Sec.~\ref{Sec:expVerification}. 

\subsection{Equation of Motion of the System}
\label{Sec. modelling_EOM}
Rather than analyzing a time-varying magnetic field, we adopt a non-inertial rotating frame in which the rotor is stationary. This choice is justified by our observation that the floater precesses with the same angular velocity as the rotor during stable levitation. We define the origin as the center of mass of the rotor. To describe the floater’s position and orientation, we use three translational coordinates to denote floater's center of mass: $x$, $y$ and $z$, along with three rotation angles: $\phi_X$,$\phi_Y$, and $\psi$ (see Appendix~\ref{appA.EOM} for further details). The Lagrangian of the floater reads 
\begin{equation}
    \mathcal{L} =\frac{1}{2} \boldsymbol{\omega} \cdot \mathbf{L} 
+ m \, \dot{\mathbf{r}} \cdot (\boldsymbol{\Omega} \times \mathbf{r}) 
+ \frac{1}{2} m \, \dot{\mathbf{r}}^2 - {U_{e}}
\end{equation}
where $\mathbf{L} =  \begin{bmatrix}I_{12} \, \omega_1 & I_{12} \, \omega_2 &
I_3 \, \omega_3 \end{bmatrix}^{\mathrm{T}}$ 
is the angular momentum of the floater, and the potential energy is given by
\begin{equation}
    {U_{e}} =mgz(t) - \mathbf{m_f} \cdot B(\mathbf{r}, \mathbf{m_r}) - \frac{1}{2} m (\boldsymbol{\Omega} \times \mathbf{r}) \cdot (\boldsymbol{\Omega} \times \mathbf{r}),
\end{equation}
with $B(\mathbf{r}, \mathbf{m_r}) = { \mu_0 (3 (\mathbf{r} \cdot \mathbf{m_r}) \cdot \mathbf{r} - \mathbf{m_r} \cdot (\mathbf{r} \cdot \mathbf{r})) }/
\left({ 4\pi \|\mathbf{r}\|^5 }\right)$. Here, we have treated the rotor and floater as two magnetic dipoles. We then apply Euler-Lagrange Equation
\begin{equation}
 \frac{\partial }{\partial t} \frac{\partial \mathcal{L}}{\partial \dot{q}} - \frac{\partial \mathcal{L}}{\partial q}=f_{d},  
\end{equation}
where $q$ is a generalized coordinate and $f_d$ dissipative force. Because ${\partial\mathcal{L}}/{\partial \psi}=0$, we have $m_3={\partial \mathcal{L}}/{\partial \dot{\psi}}$ being conserved where 
$m_3 = I_3 (\omega_r \cos \phi_x(0) \cos \phi_y(0) + \dot{\psi}(0) -\dot{\phi}_x(0) \sin \phi_y(0))$ is defined by the initial condition. By setting initial condition as stationary, $m_3=I_3 \omega_r$. We can then apply the Routh transform to eliminate the cyclic coordinate $\psi$. With $R = \mathcal{L} - m_3 \dot{\psi}(t)$, the equation of motion thus becomes 

\begin{equation}
\begin{aligned}
\left\{
\begin{array}{rl}
    \displaystyle \frac{\partial }{\partial t} \frac{\partial R}{\partial \dot{x}} - \frac{\partial R}{\partial x} 
    &= -\alpha \left(\dot{x} - \omega_r y\right); \\[2ex]
    \displaystyle \frac{\partial }{\partial t} \frac{\partial R}{\partial \dot{y}} - \frac{\partial R}{\partial y} 
    &= -\alpha \left(\omega_r x + \dot{y}\right); \\[2ex]
    \displaystyle \frac{\partial }{\partial t} \frac{\partial R}{\partial \dot{z}} - \frac{\partial R}{\partial z} 
    &= -\beta \dot{z}; \\[2ex]
    \displaystyle \frac{\partial }{\partial t} \frac{\partial R}{\partial \dot{\phi}_x} - \frac{\partial R}{\partial \phi_x} 
    &= 0; \\[2ex]
    \displaystyle \frac{\partial }{\partial t} \frac{\partial R}{\partial \dot{\phi}_y} - \frac{\partial R}{\partial \phi_y} 
    &= 0;
\end{array}
\right.
\end{aligned}
\end{equation}where the translational damping coefficient produced by the copper board is $\alpha$, the vertical $beta$ and rotational damping coefficients being ignored (see Appendix~\ref{appB.Damping} for further details). We can also approximate $I_3$ with the $2I_{12}$ (the shape of floater is an oblate cylinder) in the following analysis.

\subsection{Balance Position}
To solve for the balance position, we first set the floater as stationary and substitute lowercase letters with capital ones to denote stationary balance position rather than functions of time. We also note that higher-order terms involving $X, Y, \phi_X, \phi_Y$ can be neglected, because these are observed to be small quantities according to experimental observations. By defining $\mu=m_f m_f\mu_0/4\pi$, the balance position satisfies the following equations

\begin{equation}
\begin{cases}
m \, \omega_r^2 X + \alpha \, \omega_r \, Y + {3\mu}/{Z^4} = 0 \\
\alpha \, \omega_r \, X = m \, \omega_r^2 Y \\
m g + 3\mu (4 X - \phi_Y Z)/Z^5 = 0 \\
\phi_X=0\\
I_{12} \, \omega_r^2  \, \phi_Y   = -{\mu}/{Z^3}
\end{cases}
\label{eq:forceBalanceEquation}
\end{equation}
where magnetic forces are ${3\mu}/{Z^4}, 3\mu (4 X - \phi_Y Z)/Z^5$, magnetic torque is ${\mu}/{Z^3}$, centrifugal forces are $m \, \omega_r^2 X, m \, \omega_r^2 Y$ and damping forces $\alpha \, \omega_r \,Y$ and $ \alpha \, \omega_r \, X$.

\begin{figure}
    \centering
    \includegraphics[width=0.8\linewidth]{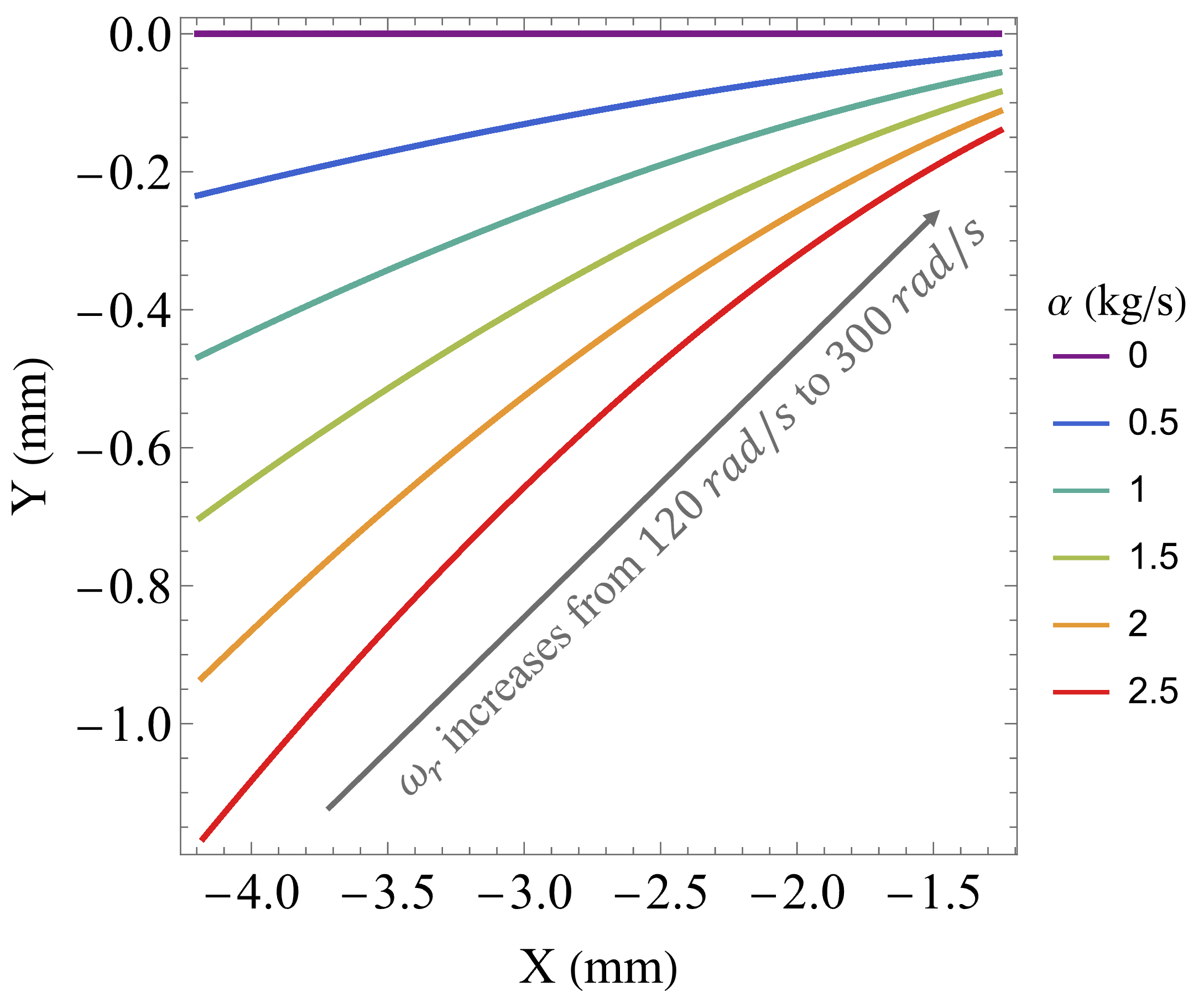}
    \caption{Stable horizontal position $(X,Y)$ as $\alpha$ and $\omega_r$ changes}
    \label{fig:XY_traj_thy}
\end{figure}

We note that these results differ from scaling laws obtained by previous studies ~\cite{le2024_stable} primarily in two perspectives. Firstly, our rotor magnet is not tilted, as explained in Sec.~\ref{Sec. modelling_EOM} Second, previous studies set small-value $X$ and $Y$ as 0, but we see here that their non-zero values actually balance horizontal magnetic forces and non-inertial forces in the rotating frame. We thus extend the calculation of the balance position and enable further stability analysis around the balance point by including horizontal position $X$ and $Y$. The stable horizontal positions $(X,Y)$ as $\alpha$ and $\omega_r$ change are shown in Fig.~\ref{fig:XY_traj_thy}. From the purple line, we see that when $\alpha = 0$, $Y$ = 0. If we also assume  $X \ll \phi_Y Z$,  it is thus possible to get the tidy analytical result 

\begin{equation}
\begin{cases}
X = -\left( {3^3\, I_{12}^4\, g^4}/{\mu\, m^3\, \omega_r^6} \right)^{1/7} \\
Z = \left( {3\, \mu^2}/{m\, g\, I_{12}\,\omega_r^2} \right)^{1/7} \\
\phi Y = -\left( {\mu \, m^3 \, g^3}/{3^3\, I_{12}^4\, \omega_r^8} \right)^{1/7}
\end{cases}
\label{eq:forceBalanceResult}
\end{equation}

As $\alpha$ increases, $Y$ starts to deviate from zero. When $\omega_{r}$ increases, the floater’s balance position shifts closer to the origin. In other words, the rotation of the rotor tends to pull the horizontal balance position toward the rotor’s axis of rotation.

Another interesting insight comes from symmetry. The rotor magnet can rotate either counter-clockwise or clockwise, and the floater’s magnetic moment can point upward or downward, determining the signs of $\mathbf{\omega_r}$ and $\mathbf{m_f}$ respectively. In the preceding analysis, we assume by default that the rotor rotates counter-clockwise and the floater’s magnetic moment is oriented upward, resulting in a horizontal equilibrium position located in the third quadrant of the non-inertial reference frame. While changing the signs of $\mathbf{\omega_{r}}$ and $mathbf{m_f}$ does not alter the floater’s stability range, it does affect its horizontal equilibrium position. In fact, by adjusting the signs, the floater can achieve equilibrium in any of the four quadrants within the non-inertial frame.

\section{Stability Analysis}
\label{Sec.Stability Analysis}
In order to analyze the stability of the motion governed by Eq.~\ref{eq:forceBalanceEquation}, we need to view positions as superposition of stable position (denoted as capital letters) and small perturbations(denoted as lowercase letters). By ignoring high order terms involving perturbations, we obtain the following linearized equation about perturbations

\[
\mathbf{M} \, \ddot{\mathbf{p}}(t) + \mathbf{C} \, \dot{\mathbf{p}}(t) + \mathbf{K} \, \mathbf{p}(t) = \mathbf{0}
\]
where
\begin{equation*}
    \mathbf{p}(t) =
    \begin{bmatrix}x(t) \\y(t) \\z(t) \\\phi_x(t) \\\phi_y(t)\end{bmatrix}, \quad
    \mathbf{M} =
    \begin{bmatrix}
    1 & 0 & 0 & 0 & 0 \\
    0 & 1 & 0 & 0 & 0 \\
    0 & 0 & 1 & 0 & 0 \\
    0 & 0 & 0 & 1 & 0 \\
    0 & 0 & 0 & 0 & 1
    \end{bmatrix}
\end{equation*}

\begin{equation*}
    \mathbf{C} =
    \begin{bmatrix}
    \alpha/m & -2 \, \omega_r & 0 & 0 & 0 \\
    2 \, \omega_r & \alpha/m & 0 & 0 & 0 \\
    0 & 0 & \beta/m & 0 & 0 \\
    0 & 0 & 0 & 0 & 0 \\
    0 & 0 & 0 & 0 & 0
    \end{bmatrix}
\end{equation*}

\begin{widetext}
\begin{equation*}
    \mathbf{K} = 
\end{equation*}
\begin{equation*}
    \label{Eq.Mat_pos}
    \begin{bmatrix}
     -\left(\omega_r^2 - \dfrac{9\mu(5X - Z\phi_Y)}{mZ^6}\right) & -\left(\dfrac{\alpha \, \omega_r}{m} - \dfrac{15\mu Y}{mZ^6}\right) & \dfrac{12\mu}{mZ^5} & \dfrac{3\mu Y}{mZ^5} & -\dfrac{3\mu(3X - Z\phi_Y)}{mZ^5} \\
     \left(\dfrac{\alpha \, \omega_r}{m} + \dfrac{15\mu Y}{mZ^6}\right) & -\left(\omega_r^2 - \dfrac{3\mu(5X - Z\phi_Y)}{mZ^6}\right) & 0 & \dfrac{3\mu X}{mZ^5} & -\dfrac{3\mu Y}{mZ^5} \\
     \dfrac{12\mu}{mZ^5} & 0 & -\dfrac{12\mu(5X - Z\phi_Y)}{mZ^6} & 0 & -\dfrac{3\mu}{mZ^4} \\
     \dfrac{3\mu Y}{I_{12} Z^5} & \dfrac{3\mu X}{I_{12} Z^5} & 0 & \dfrac{3\mu X}{I_{12} Z^4} + \omega_r^2 & 0 \\
     -\dfrac{3\mu(3X - Z\phi_Y)}{I_{12} Z^5} & -\dfrac{3\mu Y}{I_{12} Z^5} & -\dfrac{3\mu}{I_{12} Z^4} & 0 &  \dfrac{\mu(3X - Z\phi_Y)}{I_{12} Z^4} +\omega_r^2
    \end{bmatrix}
\end{equation*}
\end{widetext}from which the stability condition can be arrived by inspecting the eigenvalues $\lambda$ of the linearized system ~\cite{merkin2012_introduction}. Due to the broken symmetry introduced by the rotor’s horizontal magnetic moment and the floater’s six degrees of freedom, the stability condition, unfortunately, cannot be expressed analytically. Instead, we determine stability by numerically calculate eigenvalues: if $\text{Re}(\lambda) < 0$ for all eigenvalues, the levitation is stable.

\subsection{Visualization of Eigenvalues}
\label{Sec.StabAna_Fre}
\begin{figure}
    \makebox[\textwidth][c]{\includegraphics[width=\linewidth]{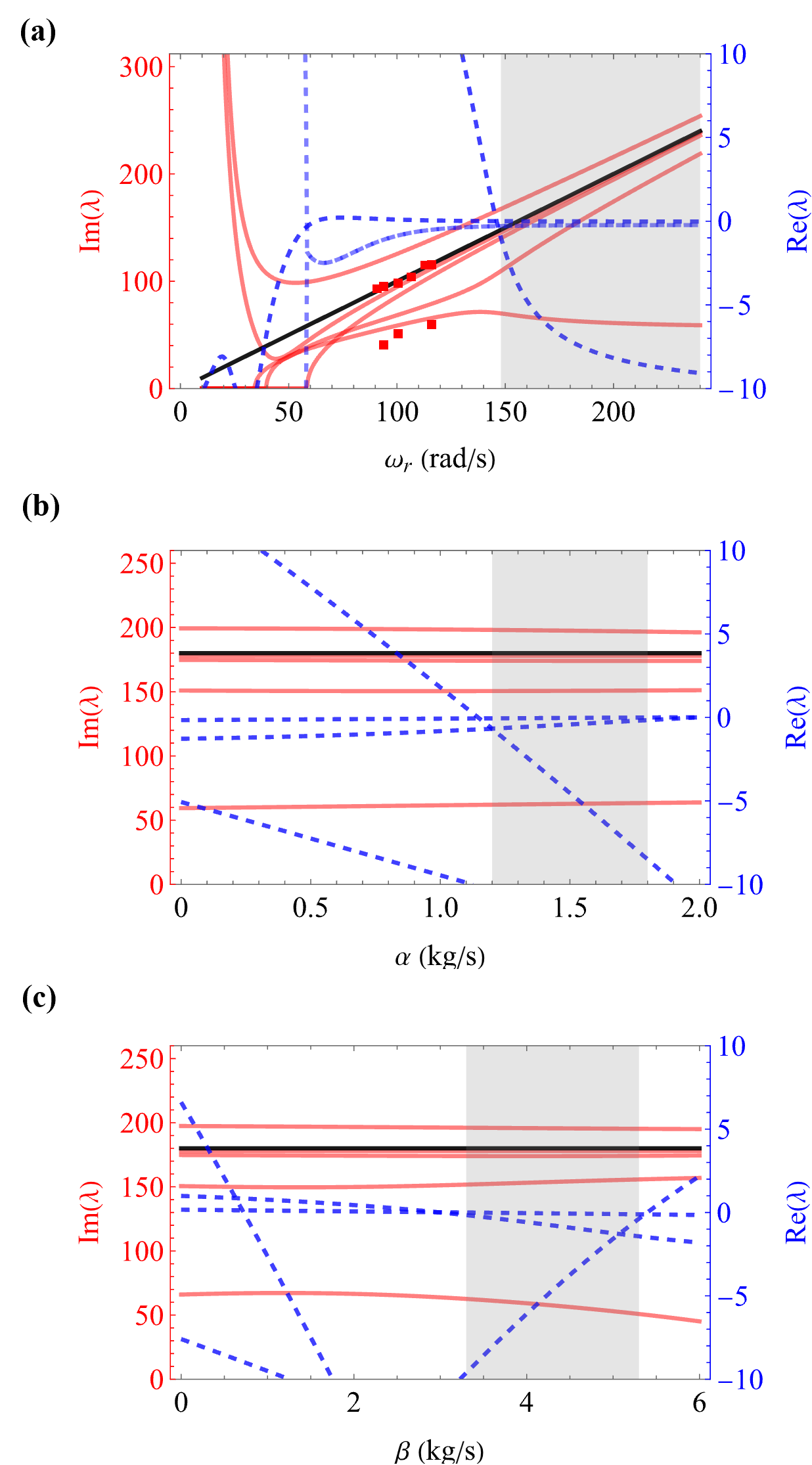}
    }
    \caption{Imaginary and real part of eigenvalues of linearized equation as $\omega_r$, $\alpha$ and $\beta$ change. The gray area denotes range of $\omega_r$, $\alpha$ or $\beta$ where for all $\lambda$, $\text{Re}(\lambda) < 0$, i.e. stable regions. In (a), red dots represents frequency peak after FFT of $x(t)$, the $x$ coordinate of center of mass, during stable levitation. }
    \label{fig:eigenVisualization}
\end{figure}

By varying $\omega_{r}$, $\alpha$, and $\beta$ numerically and using gray areas to denote stable regions, as shown in Fig.~\ref{fig:eigenVisualization}, we observe the following: i) For fixed $(\alpha, \beta)$, levitation requires $\omega_{r}$ above a threshold. ii) For fixed $(\omega_{r}, \beta)$, stability occurs within a range of $\alpha$. iii) For fixed $(\omega_{r}, \alpha)$, stability occurs within a range of $\beta$ 

The imaginary parts of the eigenvalues correspond to the system’s natural frequencies. In most cases, 4 out of 5 eigen-frequencies are close to $\omega$. However, due to the fact that Re($\lambda$) $\neq$ 0, these oscillations decay, and the system will ultimately settles at its equilibrium.

\subsection{Stability Diagrams with Parametric Space $(\omega_{r}, \alpha, \beta)$}
\label{Sec.StabAna_StabDiag}
\begin{figure}
    \centering
    \includegraphics[width=1\linewidth]{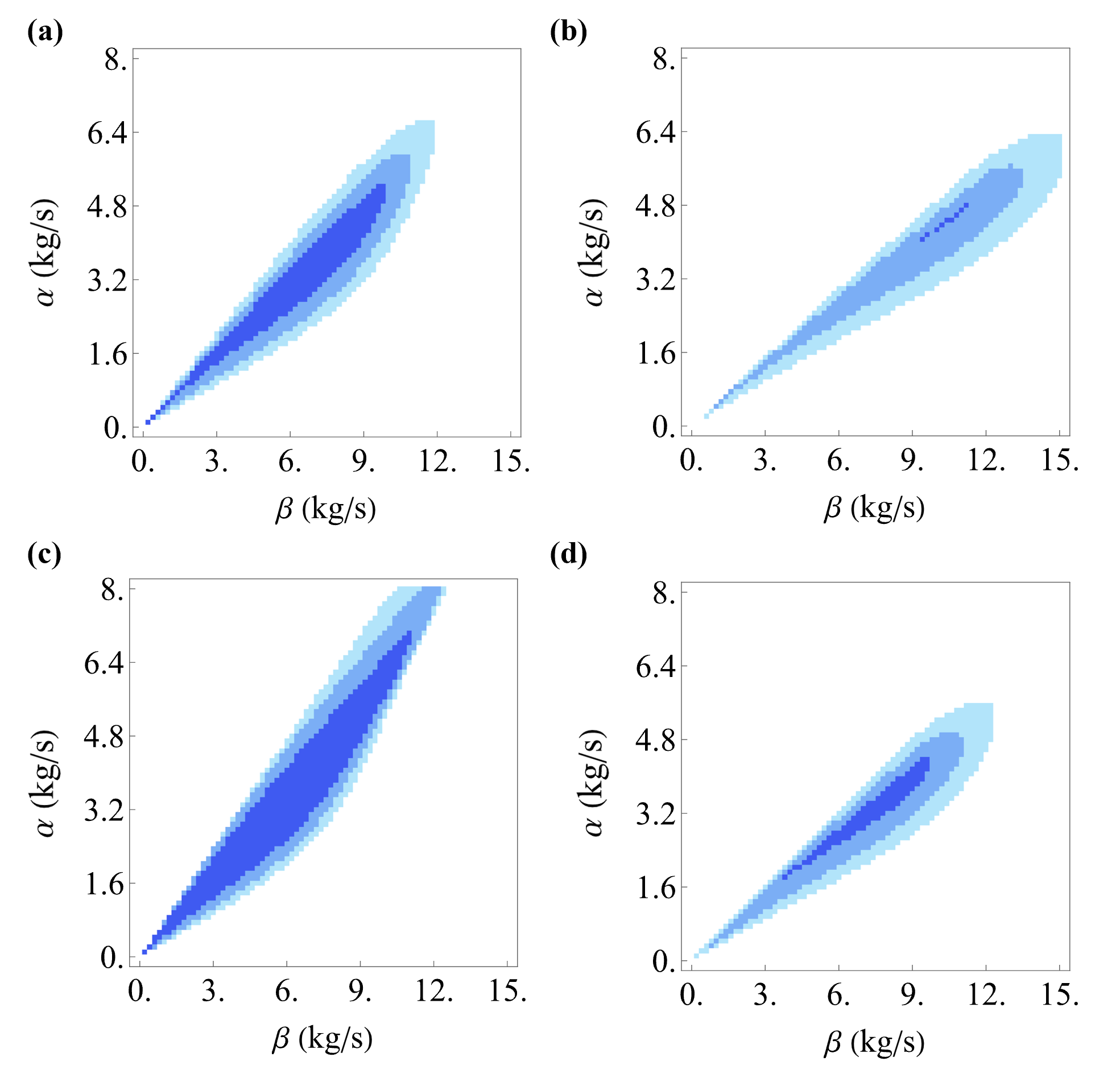}
    \caption{Stability diagram with parametric space $\left( \alpha,\beta \right)$. Blue regions demonstrate stability region of $\omega_r = $160, 190, 220 $\mathrm{rad/s}$ respectively as the color becomes lighter and lighter. The geometric and magnetic parameters in (a) are 
    $I_{12}=1.2 \times 10^{-5} \mathrm{kg\cdot \, m^2}$, 
    $m_f=6.5 \,\mathrm{A\cdot m^2}$,
    $m_f=3.6\, \mathrm{A\cdot m^2}$ and
    $m=0.075 \, \mathrm{kg}$.
    (b) changes $m$ to $1.2m$ with other parameters unchanged from (a). 
    (c) changes $I_{12}$ to $1.2I_{12}$ with other parameters unchanged from (a). 
    (d) changes $m_f$ to $1.5m_f$ with other parameters unchanged from (a).
     }
    \label{fig:stabilityRegion_a_b}
\end{figure}

After traversing the values of $\alpha$, $\beta$, and $\omega_{r}$, we can also obtain the stability phase diagrams plotted in the $(\alpha, \beta)$ parametric space, as illustrated in Fig.~\ref{fig:stabilityRegion_a_b}. Here, $\alpha$, $\beta$ have a range of the same order as our measurement of the copper board that enables stable levitation (see Appendix~\ref{appB.Damping} for details of measurement). In Fig.~\ref{fig:stabilityRegion_a_b} (a), we can see that the stability region is bounded within the $(\alpha, \beta)$ parametric space, and increases its area when $\omega_r$ increases. The bounded nature also suggests a counterintuitive result: stability cannot be attributed solely to the presence of large damping forces. In fact, stable levitation is achievable even at relatively small values of both $\alpha$ and $\beta$. It is also important to note that in most practical scenarios, the minimum damping coefficients required for stability significantly exceed those of a floater moving in air. Consequently, in this particular system, achieving stable levitation typically necessitates additional damping sources, such as a conductive metal or a viscous fluid.

We also note that our theoretical model qualitatively agrees with the experimental observations reported in previous study ~\cite{hermansen2023_rot}, particularly in describing how the minimum $\omega_r$ required for stable levitation depends on the geometric and magnetic parameters of the system. It was reported that smaller floater magnets require higher $\omega_r$, consistent with our theoretical prediction that larger value of $I_{12}$ corresponds to a wider stability region, as illustrated by the comparison between Fig.~\ref{fig:stabilityRegion_a_b} (a) and (c). Furthermore, the experimental observation that a larger magnetic moment of the rotor magnet does not necessarily facilitate stable levitation aligns with the theoretical finding that increasing $m_f$ may reduce the stability region in the $(\alpha,\beta)$ parametric space, as illustrated by the comparison between Fig.~\ref{fig:stabilityRegion_a_b} (a) and (d). 

\section{Experimental Results}
\label{Sec:expVerification}
\begin{figure}
    \centering
    \includegraphics[width=0.9\textwidth]{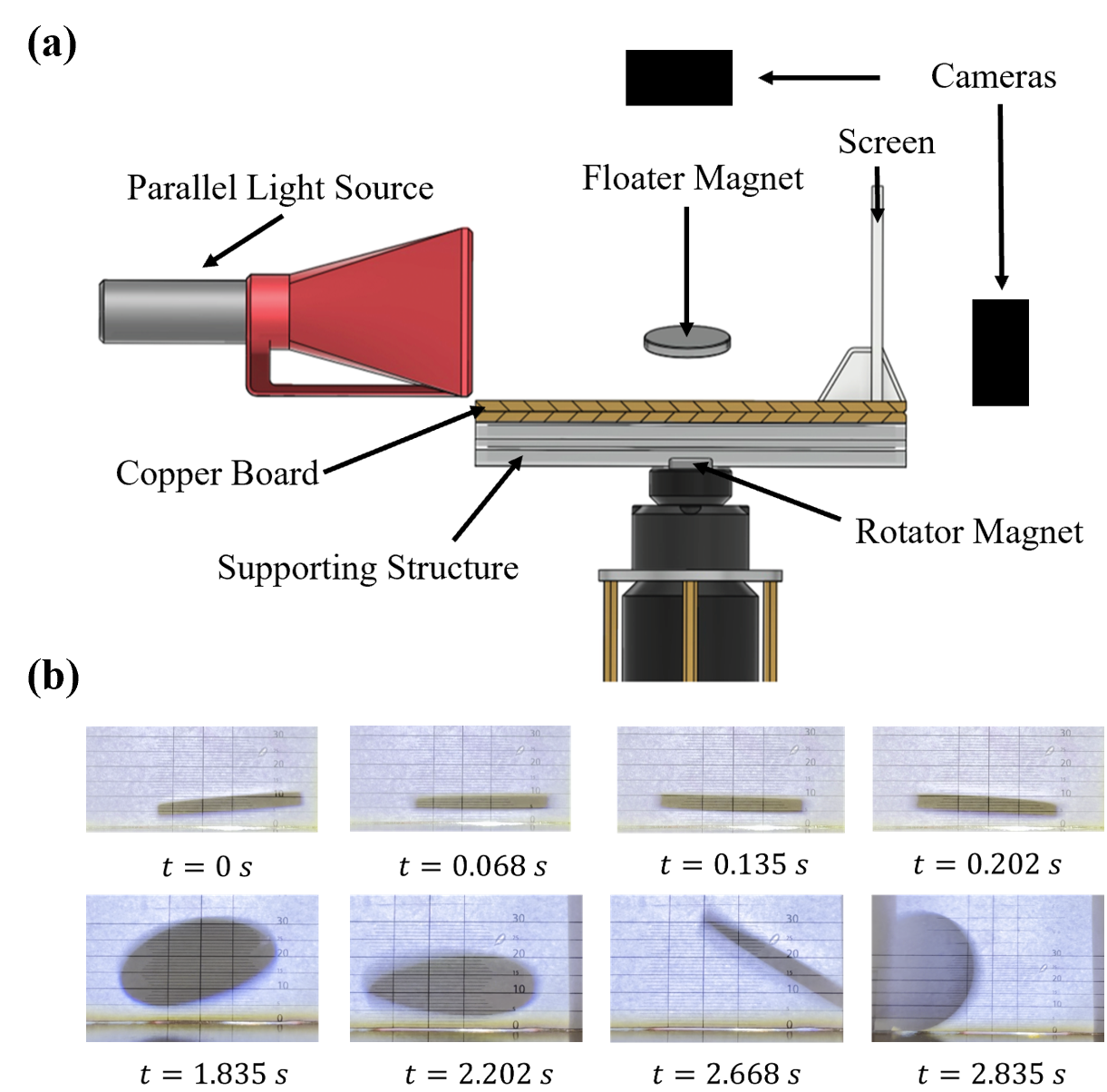}
    \caption{
    (a) Side view of the system's configuration. The parallel light source, primarily consisting of a Fresnel lens, along with the screen with scale marks, enables accurate measurement of stability height.
    (b) Side view of the levitating floater from the perspective of the screen. The first row demonstrates a typical stable levitation, with the second unstable. }
    \label{fig:config}
\end{figure}

In order to compare our theoretical results with experiments, we need to record the position of the floater as the rotational speed of the rotor changes. Two cameras, placed on the side and top of the apparatus, were used to achieve this, as shown in Fig.~\ref{fig:config}. As scaling laws of balance position similar to Eq.~\ref{eq:forceBalanceResult} have been quantitatively verified in previous studies~\cite{le2024_stable}, our emphasis in the following sections will then be on stability boundary in parametric space and natural frequencies of oscillation around balance position during stable levitation. 

\subsection{Stability Diagram}
\begin{figure}
    \centering
    \includegraphics[width=1\linewidth]{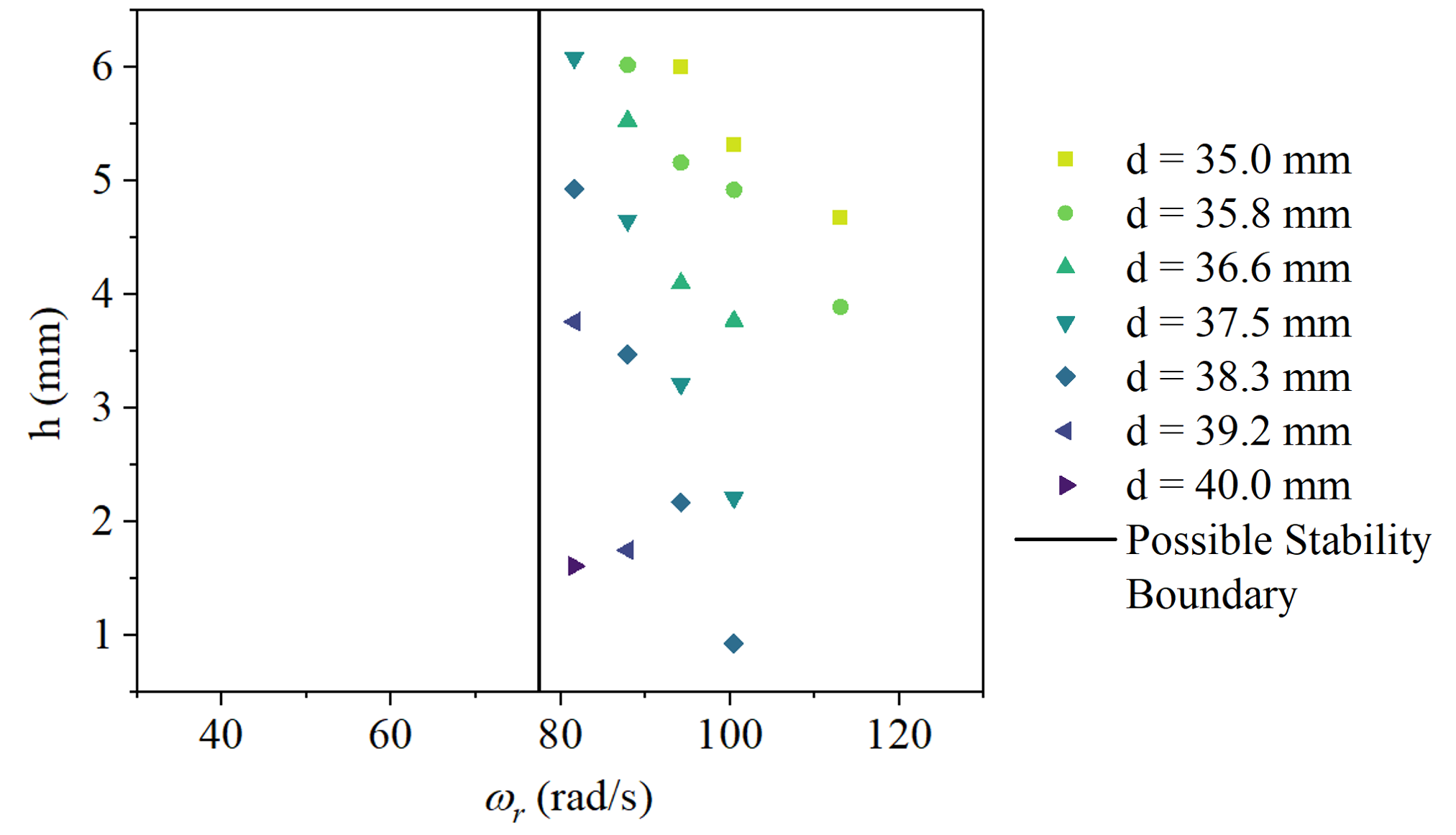}
    \caption{Experimental Stability Diagram. The dots represent collected $\omega_r$ and $h$ during stable levitation near the boundary between stable and unstable motion, leading to a rough experimental boundary between stable levitation and unstable status, denoted by the black line. }
    \label{fig:stabilityRegion_h_wr}
\end{figure}

Given a specific set of rotor, floater magnets and copper board, the three dimensional parametric space $(\alpha,\beta,\omega_r)$ of the system can be reduced into two dimensional $(h,\omega_r)$, because $\alpha$ and $\beta$ are coupled through the distance between copper board and floater magnet, termed $h$. Note that the floating height $z$ is the sum of $h$ and $d$, the latter being the distance between rotor magnet and copper board, as defined in Fig.~\ref{fig:coordinateDef}. By adjusting $d$, measuring $h$ in stable levitation, and reducing the angular velocity of rotor $\omega_r$ until the floater magnets fail to levitate stably, we can plot several sets of $(h,\omega_r)$ near the stability boundary, as shown in Fig.~\ref{fig:stabilityRegion_h_wr}. Here, in experiment, we define stable levitation as the floater levitating with non-increasing amplitude of oscillation for at least 30 s. 

We also note an interesting experimental result: increasing $d$—which decreases $h$ and thereby increases both $\alpha$ and $\beta$—does not necessary impliy lower minimum $\omega_r$ required for stable levitation. This behavior stems from the bounded nature of the stability region in the $(\alpha, \beta)$ plane at fixed $\omega_{r}$, as discussed in Sec.~\ref{Sec.StabAna_StabDiag}.

\subsection{Motion During Stable Levitation}
\begin{figure}
    \centering
    \includegraphics[width=0.85\textwidth]{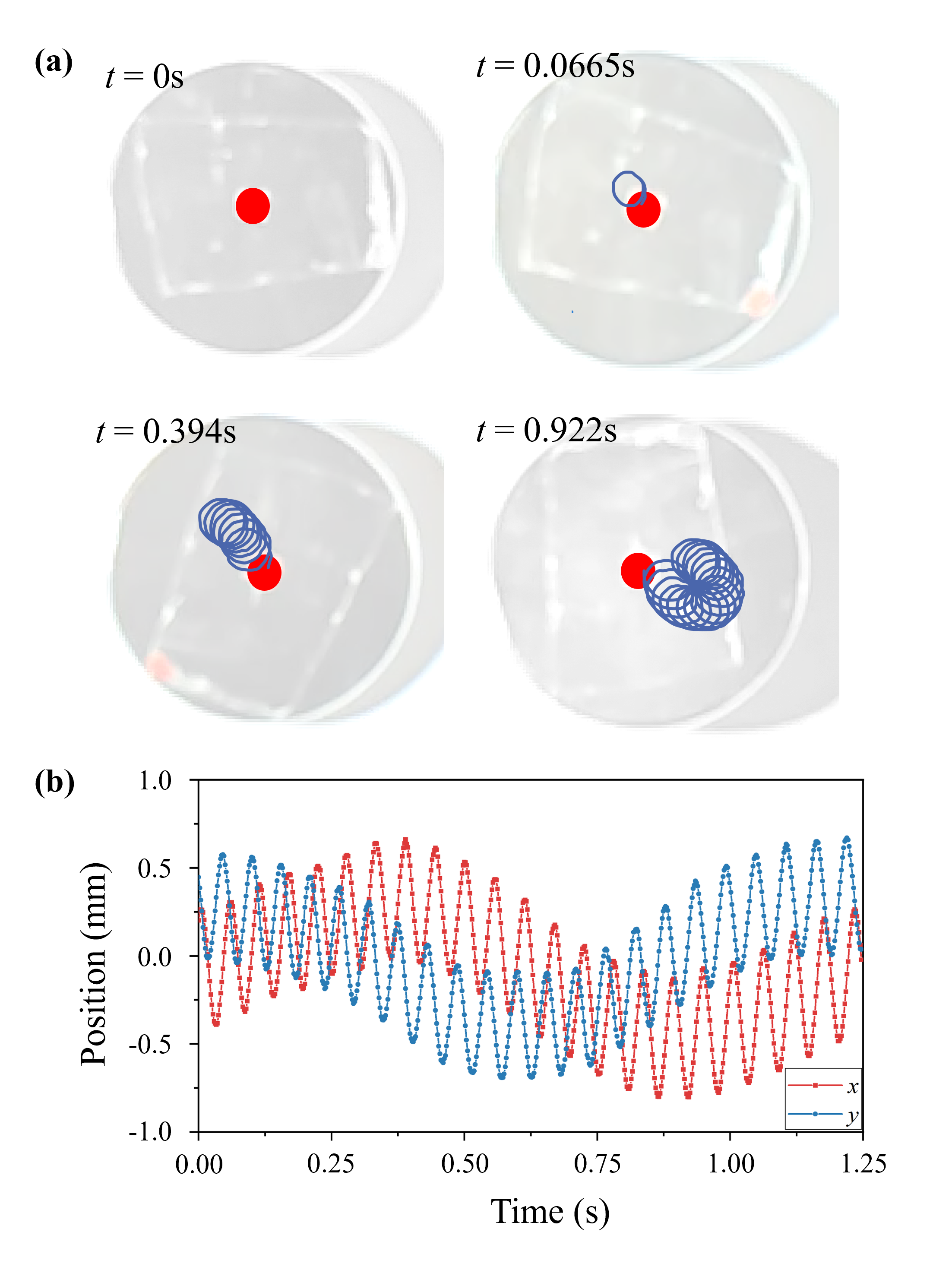}
    \caption{Trajectories of the center of mass from top views.}
    \label{fig:combined_trajectory}
\end{figure}

During levitation, we observed that the floater oscillates in the horizontal ($x$ and $y$) directions, resulting in trajectories reminiscent of those traced by a heavy spinning top, as illustrated in Fig.~\ref{fig:combined_trajectory}. The $x$ and $y$ displacements appear to be composed of a superposition of at least two sinusoidal components, one of which is approximately the driving frequency $\omega_{r}$. We plotted the frequency peaks of FFT of $x(t)$ during stable levitation, as shown by the red dots in Fig.~\ref{fig:eigenVisualization} (a). From the graph, we can see that the theoretical prediction of natural frequencies of motion closely matches the experimental results.

\section{Conclusion}
\label{Sec.conclusion}
In this study, we present a comprehensive analysis of stable magnetic levitation achieved by a floater magnet above a rotating magnet and copper board system. Our results demonstrate that stable levitation occurs only within well-defined ranges of rotation speed and damping coefficients, with the copper board’s damping effect playing a critical stabilizing role. Combining experimental observations with theoretical modeling, we reveal intricate dynamic behaviors, including frequency coupling, and construct a detailed stability phase diagram that elucidates the interplay between magnetic forces, damping, and rotational dynamics. Beyond its fundamental insights, this work paves the way for practical applications, such as precision manipulation of ferromagnetic particles, as well as educational demonstrations of stability in rotating magnetic systems.

\acknowledgments{The authors thank Houxi Zhou for helpful discussion. This work is supported by the National Natural Science Foundation of China under grant No. 12305037 and the Fundamental Research Funds for the Central Universities under grant No. 2023NTST017.}

\appendix

\begin{figure*}
\includegraphics[width=1\linewidth]{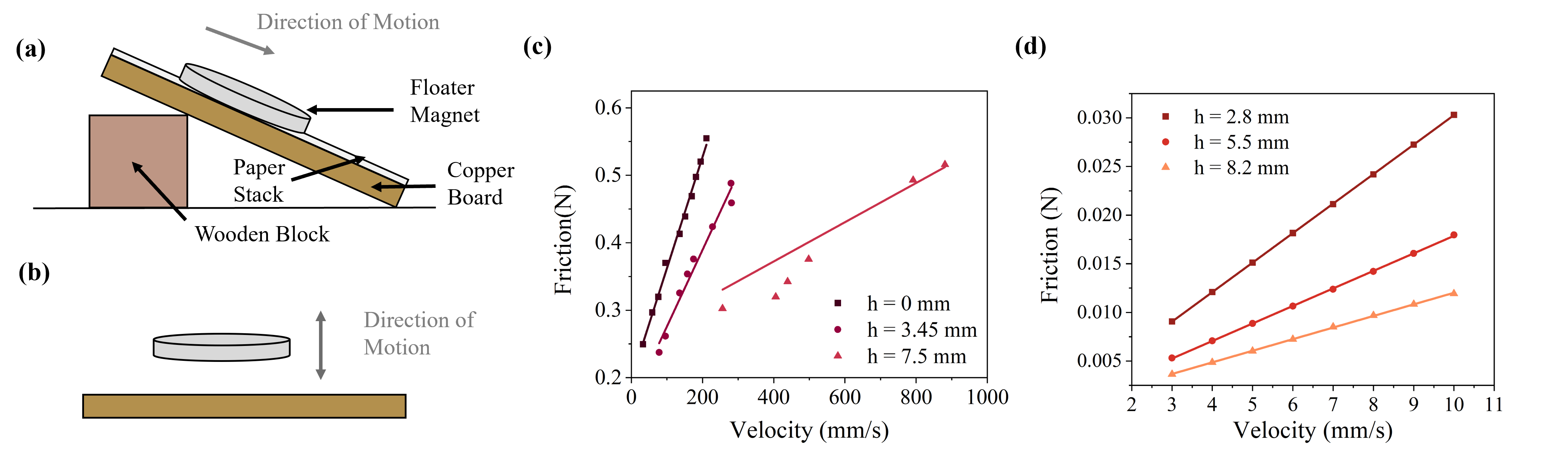}
\caption{\label{fig:Damping_Measurement}(a) Sketch of measurement of horizontal damping coefficient (b) Sketch of measurement of vertical damping coefficient. (c) Horizontal friction-velocity linear relationship. The gradient of each line is fitted to obtain $\alpha(h)$ (d) Vertical friction-velocity linear relationship. The gradient of each line is fitted to obtain $\beta(h)$}
\end{figure*}

\section{Definition of Coordinate System}
\label{appA.EOM}

\begin{figure}
    \centering
    \includegraphics[width=0.75\linewidth]{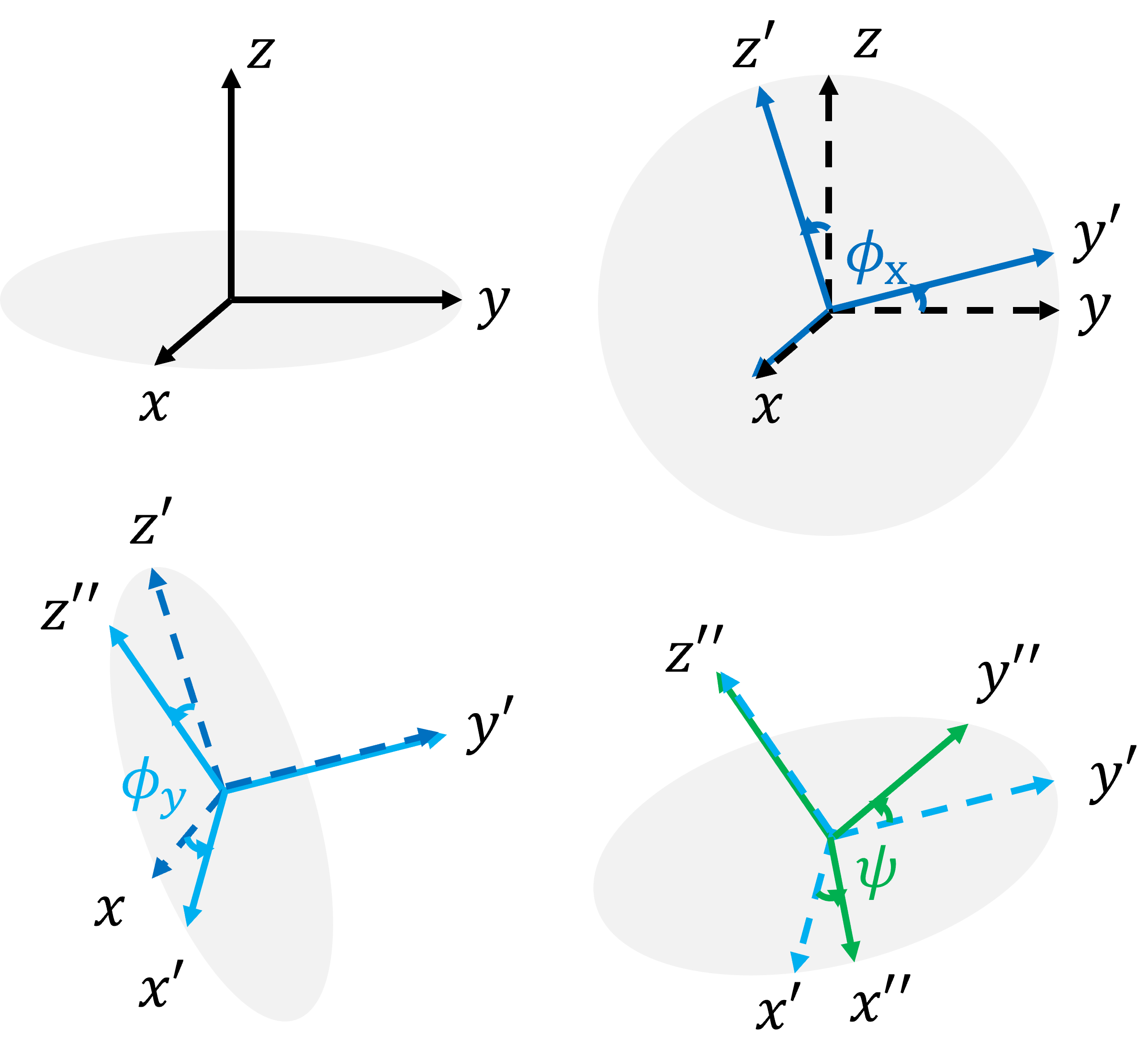}
    \caption{$\phi_X$ is rotational angle about the $x$-axis, $\phi_Y$ about the rotated $y'$-axis, and $\psi$ about $z''$-axis}
    \label{fig:RotCoor}
\end{figure}

According to the definition of our system's coordinate, which is shown in Fig.~\ref{fig:RotCoor}, the angular velocity of the floater in its body frame reads
\begin{equation}
    \boldsymbol{\omega} = 
    \mathbf{R}_3 \mathbf{R}_2 \mathbf{R}_1 \begin{bmatrix}0 \\ 0 \\ \omega_r \end{bmatrix}
    + \mathbf{R}_3 \mathbf{R}_2 \begin{bmatrix}\dot{\phi}_x(t) \\ 0 \\ 0 \end{bmatrix}
    + \mathbf{R}_3 \begin{bmatrix}0 \\ \dot{\phi}_y(t) \\0\end{bmatrix}
    + \begin{bmatrix}0 \\0 \\\dot{\psi}(t)\end{bmatrix}
\end{equation} where

\begin{align}
    \mathbf{R_1} &=
    \begin{pmatrix}
        1 & 0 & 0 \\
        0 & \cos \phi_x(t) & \sin \phi_x(t) \\
        0 & -\sin \phi_x(t) & \cos \phi_x(t)
    \end{pmatrix} \label{eq:R1} \\
    \mathbf{R_2} &=
    \begin{pmatrix}
        \cos \phi_y(t) & 0 & -\sin \phi_y(t) \\
        0 & 1 & 0 \\
        \sin \phi_y(t) & 0 & \cos \phi_y(t)
    \end{pmatrix} \label{eq:R2} \\
    \mathbf{R_3} &=
    \begin{pmatrix}
        \cos \psi(t) & \sin \psi(t) & 0 \\
        -\sin \psi(t) & \cos \psi(t) & 0 \\
        0 & 0 & 1
    \end{pmatrix} \label{eq:R3}
\end{align}

Using these rotational matrices, the postures of floater and rotor can also be expressed respectively as
\begin{align}
    \mathbf{m_f} &= m_f \cdot \mathbf{R}_1^{-1} \cdot \mathbf{R}_2^{-1} \cdot \mathbf{R}_3^{-1} \cdot 
    \left[\begin{array}{ccc} 0 & 0 & 1 \end{array}\right]^T \\
    \mathbf{m_r} &= m_f \cdot
    \left[\begin{array}{ccc} 1 & 0 & 0 \end{array}\right]^T
\end{align}

\section{Effect of Copper Board and Measurement of Damping Coefficients}
\label{appB.Damping}
The copper board plays a crucial role in achieving stable levitation. Replacing it with a non-conductive material, such as wood, eliminates levitation under identical $\omega_{r}$ conditions. However, introducing a viscous fluid reinstates levitation, suggesting that damping is a critical factor in the system’s stability. Given the complex interactions among the floater, the rotor, and the copper board, we begin by examining the role of the copper board in detail before modeling it as a simplified damping element. The copper board fulfills two primary functions. First, it provides inductive damping, opposing the motion of both the rotor and the floater. Second, it imposes a physical lower bound on the levitation height. Since the second function lies beyond the scope of our stability analysis, we restrict our focus to inductive damping. To maintain analytical tractability, we consider the board’s interaction with the rotor and the floater independently, rather than as a coupled system.

In the absence of the floater, the motion of the rotor remains effectively unaltered due to its externally maintained angular velocity. Our measurements indicate that the variation in the magnetic field caused by the presence of the copper board is $<5\%$ in both vertical and horizontal directions. Consequently, this effect is neglected in our analysis. Conversely, in the absence of the rotor, the floater alone experiences damping due to eddy currents induced in the copper board. In our simplified model, we consider only translational damping, which includes both vertical and horizontal components, characterized by the coefficients $\beta(h)$ and $\alpha(h)$, respectively, where $h$ denotes the vertical distance between the floater and the copper board.

Aiming at arriving $f_{horizontal} =\alpha(h)\,v$ and $f_{vertical} = \beta(h)\, v$, we need to design experiments in which all $v$, $h$, and $f$ are measurable. The experimental operations are depicted in following subsections. After that, alpha and beta can be arrived by linearly fitting $f-v$ curve for each $h$. Then we can arrive at $\alpha(h)$ and $\beta(h)$ by interpolation function. 

\subsection{Horizontal Damping Coefficient}
In order to measure the damping coefficient in horizontal direction, we slant the copper board, place the floater magnet on top of it, and let the floater magnet to naturally slide down. Because the floater slides in uniform velocity, the friction can be calculated by force balance. Adjusting the slant angle, the constant velocity of floater changes; adding paper between the copper board and floater, we can change the distance between them, as shown in Fig.~\ref{fig:Damping_Measurement} (a) and (c). Thus all $v$, $h$, and $f$ are measurable. 

\subsection{Vertical Damping Coefficient}
Unlike the horizontal case, measuring the vertical damping coefficient is more challenging. One method involves moving the floater vertically at constant speed and measuring the resulting damping force. However, this requires precise control and sensing equipment beyond our current capabilities. As an alternative, we implement this procedure in Comsol Multiphysics and measure $v$, $h$, and $f$ as floater moves in vertical direction, as shown in Fig.~\ref{fig:Damping_Measurement} (b) and (d).

\FloatBarrier
\bibliography{apssamp}

\begin{thebibliography}{24}%
\makeatletter
\providecommand \@ifxundefined [1]{%
 \@ifx{#1\undefined}
}%
\providecommand \@ifnum [1]{%
 \ifnum #1\expandafter \@firstoftwo
 \else \expandafter \@secondoftwo
 \fi
}%
\providecommand \@ifx [1]{%
 \ifx #1\expandafter \@firstoftwo
 \else \expandafter \@secondoftwo
 \fi
}%
\providecommand \natexlab [1]{#1}%
\providecommand \enquote  [1]{``#1''}%
\providecommand \bibnamefont  [1]{#1}%
\providecommand \bibfnamefont [1]{#1}%
\providecommand \citenamefont [1]{#1}%
\providecommand \href@noop [0]{\@secondoftwo}%
\providecommand \href [0]{\begingroup \@sanitize@url \@href}%
\providecommand \@href[1]{\@@startlink{#1}\@@href}%
\providecommand \@@href[1]{\endgroup#1\@@endlink}%
\providecommand \@sanitize@url [0]{\catcode `\\12\catcode `\$12\catcode `\&12\catcode `\#12\catcode `\^12\catcode `\_12\catcode `\%12\relax}%
\providecommand \@@startlink[1]{}%
\providecommand \@@endlink[0]{}%
\providecommand \url  [0]{\begingroup\@sanitize@url \@url }%
\providecommand \@url [1]{\endgroup\@href {#1}{\urlprefix }}%
\providecommand \urlprefix  [0]{URL }%
\providecommand \Eprint [0]{\href }%
\providecommand \doibase [0]{https://doi.org/}%
\providecommand \selectlanguage [0]{\@gobble}%
\providecommand \bibinfo  [0]{\@secondoftwo}%
\providecommand \bibfield  [0]{\@secondoftwo}%
\providecommand \translation [1]{[#1]}%
\providecommand \BibitemOpen [0]{}%
\providecommand \bibitemStop [0]{}%
\providecommand \bibitemNoStop [0]{.\EOS\space}%
\providecommand \EOS [0]{\spacefactor3000\relax}%
\providecommand \BibitemShut  [1]{\csname bibitem#1\endcsname}%
\let\auto@bib@innerbib\@empty
\bibitem [{\citenamefont {Turker}\ and\ \citenamefont {Arslan-Yildiz}(2018)}]{turker2018_magLev1}%
  \BibitemOpen
  \bibfield  {author} {\bibinfo {author} {\bibfnamefont {E.}~\bibnamefont {Turker}}\ and\ \bibinfo {author} {\bibfnamefont {A.}~\bibnamefont {Arslan-Yildiz}},\ }\bibfield  {title} {\bibinfo {title} {Recent advances in magnetic levitation: a biological approach from diagnostics to tissue engineering},\ }\href {https://doi.org/https://doi.org/10.1021/acsbiomaterials.7b00700} {\bibfield  {journal} {\bibinfo  {journal} {ACS Biomaterials Science \& Engineering}\ }\textbf {\bibinfo {volume} {4}},\ \bibinfo {pages} {787} (\bibinfo {year} {2018})}\BibitemShut {NoStop}%
\bibitem [{\citenamefont {Ge}\ \emph {et~al.}(2020{\natexlab{a}})\citenamefont {Ge}, \citenamefont {Nemiroski}, \citenamefont {Mirica}, \citenamefont {Mace}, \citenamefont {Hennek}, \citenamefont {Kumar},\ and\ \citenamefont {Whitesides}}]{ge2020_magLev2}%
  \BibitemOpen
  \bibfield  {author} {\bibinfo {author} {\bibfnamefont {S.}~\bibnamefont {Ge}}, \bibinfo {author} {\bibfnamefont {A.}~\bibnamefont {Nemiroski}}, \bibinfo {author} {\bibfnamefont {K.~A.}\ \bibnamefont {Mirica}}, \bibinfo {author} {\bibfnamefont {C.~R.}\ \bibnamefont {Mace}}, \bibinfo {author} {\bibfnamefont {J.~W.}\ \bibnamefont {Hennek}}, \bibinfo {author} {\bibfnamefont {A.~A.}\ \bibnamefont {Kumar}},\ and\ \bibinfo {author} {\bibfnamefont {G.~M.}\ \bibnamefont {Whitesides}},\ }\bibfield  {title} {\bibinfo {title} {Magnetic levitation in chemistry, materials science, and biochemistry},\ }\href {https://doi.org/https://doi.org/10.1002/anie.201903391} {\bibfield  {journal} {\bibinfo  {journal} {Angewandte Chemie International Edition}\ }\textbf {\bibinfo {volume} {59}},\ \bibinfo {pages} {17810} (\bibinfo {year} {2020}{\natexlab{a}})}\BibitemShut {NoStop}%
\bibitem [{\citenamefont {Ge}\ \emph {et~al.}(2020{\natexlab{b}})\citenamefont {Ge}, \citenamefont {Nemiroski}, \citenamefont {Mirica}, \citenamefont {Mace}, \citenamefont {Hennek}, \citenamefont {Kumar},\ and\ \citenamefont {Whitesides}}]{Ge2020MagLevApp}%
  \BibitemOpen
  \bibfield  {author} {\bibinfo {author} {\bibfnamefont {S.}~\bibnamefont {Ge}}, \bibinfo {author} {\bibfnamefont {A.}~\bibnamefont {Nemiroski}}, \bibinfo {author} {\bibfnamefont {K.~A.}\ \bibnamefont {Mirica}}, \bibinfo {author} {\bibfnamefont {C.~R.}\ \bibnamefont {Mace}}, \bibinfo {author} {\bibfnamefont {J.~W.}\ \bibnamefont {Hennek}}, \bibinfo {author} {\bibfnamefont {A.~A.}\ \bibnamefont {Kumar}},\ and\ \bibinfo {author} {\bibfnamefont {G.~M.}\ \bibnamefont {Whitesides}},\ }\bibfield  {title} {\bibinfo {title} {Magnetic levitation in chemistry, materials science, and biochemistry},\ }\href {https://doi.org/https://doi.org/10.1002/anie.201903391} {\bibfield  {journal} {\bibinfo  {journal} {Angewandte Chemie International Edition}\ }\textbf {\bibinfo {volume} {59}},\ \bibinfo {pages} {17810} (\bibinfo {year} {2020}{\natexlab{b}})},\ \Eprint {https://arxiv.org/abs/https://onlinelibrary.wiley.com/doi/pdf/10.1002/anie.201903391} {https://onlinelibrary.wiley.com/doi/pdf/10.1002/anie.201903391} \BibitemShut
  {NoStop}%
\bibitem [{\citenamefont {Simon}\ \emph {et~al.}(2001)\citenamefont {Simon}, \citenamefont {Heflinger},\ and\ \citenamefont {Geim}}]{simon2001diamag}%
  \BibitemOpen
  \bibfield  {author} {\bibinfo {author} {\bibfnamefont {M.~D.}\ \bibnamefont {Simon}}, \bibinfo {author} {\bibfnamefont {L.~O.}\ \bibnamefont {Heflinger}},\ and\ \bibinfo {author} {\bibfnamefont {A.~K.}\ \bibnamefont {Geim}},\ }\bibfield  {title} {\bibinfo {title} {Diamagnetically stabilized magnet levitation},\ }\href {https://doi.org/https://doi.org/10.1119/1.1375157} {\bibfield  {journal} {\bibinfo  {journal} {American Journal of Physics}\ }\textbf {\bibinfo {volume} {69}},\ \bibinfo {pages} {702} (\bibinfo {year} {2001})}\BibitemShut {NoStop}%
\bibitem [{\citenamefont {Supreeth}\ \emph {et~al.}(2022)\citenamefont {Supreeth}, \citenamefont {Bekinal}, \citenamefont {Chandranna},\ and\ \citenamefont {Doddamani}}]{supreeth2022_superconReview}%
  \BibitemOpen
  \bibfield  {author} {\bibinfo {author} {\bibfnamefont {D.}~\bibnamefont {Supreeth}}, \bibinfo {author} {\bibfnamefont {S.~I.}\ \bibnamefont {Bekinal}}, \bibinfo {author} {\bibfnamefont {S.~R.}\ \bibnamefont {Chandranna}},\ and\ \bibinfo {author} {\bibfnamefont {M.}~\bibnamefont {Doddamani}},\ }\bibfield  {title} {\bibinfo {title} {A review of superconducting magnetic bearings and their application},\ }\href {https://doi.org/https://doi.org/10.1109/TASC.2022.3156813} {\bibfield  {journal} {\bibinfo  {journal} {IEEE Transactions on Applied Superconductivity}\ }\textbf {\bibinfo {volume} {32}},\ \bibinfo {pages} {1} (\bibinfo {year} {2022})}\BibitemShut {NoStop}%
\bibitem [{\citenamefont {Simon}\ \emph {et~al.}(1997)\citenamefont {Simon}, \citenamefont {Heflinger},\ and\ \citenamefont {Ridgway}}]{simon1997_levitron}%
  \BibitemOpen
  \bibfield  {author} {\bibinfo {author} {\bibfnamefont {M.~D.}\ \bibnamefont {Simon}}, \bibinfo {author} {\bibfnamefont {L.~O.}\ \bibnamefont {Heflinger}},\ and\ \bibinfo {author} {\bibfnamefont {S.}~\bibnamefont {Ridgway}},\ }\bibfield  {title} {\bibinfo {title} {Spin stabilized magnetic levitation},\ }\href {https://doi.org/10.1119/1.18488} {\bibfield  {journal} {\bibinfo  {journal} {American Journal of Physics}\ }\textbf {\bibinfo {volume} {65}},\ \bibinfo {pages} {286} (\bibinfo {year} {1997})}\BibitemShut {NoStop}%
\bibitem [{\citenamefont {P{\'e}rez}\ and\ \citenamefont {Garc{\'\i}a-S{\'a}nchez}(2015)}]{perez2015_drivenLevitron}%
  \BibitemOpen
  \bibfield  {author} {\bibinfo {author} {\bibfnamefont {A.~T.}\ \bibnamefont {P{\'e}rez}}\ and\ \bibinfo {author} {\bibfnamefont {P.}~\bibnamefont {Garc{\'\i}a-S{\'a}nchez}},\ }\bibfield  {title} {\bibinfo {title} {Dynamics of a levitron under a periodic magnetic forcing},\ }\href {https://doi.org/https://doi.org/10.1119/1.4895800} {\bibfield  {journal} {\bibinfo  {journal} {American Journal of Physics}\ }\textbf {\bibinfo {volume} {83}},\ \bibinfo {pages} {133} (\bibinfo {year} {2015})}\BibitemShut {NoStop}%
\bibitem [{\citenamefont {Baldwin}\ \emph {et~al.}(2018)\citenamefont {Baldwin}, \citenamefont {de~Fouchier}, \citenamefont {Atkinson}, \citenamefont {Hill}, \citenamefont {Swift},\ and\ \citenamefont {Fairhurst}}]{baldwin2018_flea}%
  \BibitemOpen
  \bibfield  {author} {\bibinfo {author} {\bibfnamefont {K.~A.}\ \bibnamefont {Baldwin}}, \bibinfo {author} {\bibfnamefont {J.-B.}\ \bibnamefont {de~Fouchier}}, \bibinfo {author} {\bibfnamefont {P.}~\bibnamefont {Atkinson}}, \bibinfo {author} {\bibfnamefont {R.}~\bibnamefont {Hill}}, \bibinfo {author} {\bibfnamefont {M.}~\bibnamefont {Swift}},\ and\ \bibinfo {author} {\bibfnamefont {D.}~\bibnamefont {Fairhurst}},\ }\bibfield  {title} {\bibinfo {title} {Magnetic levitation stabilized by streaming fluid flows},\ }\href {https://doi.org/https://doi.org/10.1103/PhysRevLett.121.064502} {\bibfield  {journal} {\bibinfo  {journal} {Physical Review Letters}\ }\textbf {\bibinfo {volume} {121}},\ \bibinfo {pages} {064502} (\bibinfo {year} {2018})}\BibitemShut {NoStop}%
\bibitem [{\citenamefont {Perdriat}\ \emph {et~al.}(2023)\citenamefont {Perdriat}, \citenamefont {Pellet-Mary}, \citenamefont {Copie},\ and\ \citenamefont {H\'etet}}]{perdriat2023_paulTrap}%
  \BibitemOpen
  \bibfield  {author} {\bibinfo {author} {\bibfnamefont {M.}~\bibnamefont {Perdriat}}, \bibinfo {author} {\bibfnamefont {C.}~\bibnamefont {Pellet-Mary}}, \bibinfo {author} {\bibfnamefont {T.}~\bibnamefont {Copie}},\ and\ \bibinfo {author} {\bibfnamefont {G.}~\bibnamefont {H\'etet}},\ }\bibfield  {title} {\bibinfo {title} {Planar magnetic paul traps for ferromagnetic particles},\ }\href {https://doi.org/https://10.1103/PhysRevResearch.5.L032045} {\bibfield  {journal} {\bibinfo  {journal} {Phys. Rev. Res.}\ }\textbf {\bibinfo {volume} {5}},\ \bibinfo {pages} {L032045} (\bibinfo {year} {2023})}\BibitemShut {NoStop}%
\bibitem [{\citenamefont {Kirillov}(2021)}]{Kirillov2021nonConStab}%
  \BibitemOpen
  \bibfield  {author} {\bibinfo {author} {\bibfnamefont {O.~N.}\ \bibnamefont {Kirillov}},\ }\href {https://doi.org/https://doi.org/10.1515/9783110655407} {\emph {\bibinfo {title} {Nonconservative Stability Problems of Modern Physics}}}\ (\bibinfo  {publisher} {De Gruyter},\ \bibinfo {address} {Berlin, Boston},\ \bibinfo {year} {2021})\BibitemShut {NoStop}%
\bibitem [{\citenamefont {Farquhar}\ and\ \citenamefont {Kamel}(1973)}]{farquhar1973_quasi}%
  \BibitemOpen
  \bibfield  {author} {\bibinfo {author} {\bibfnamefont {R.~W.}\ \bibnamefont {Farquhar}}\ and\ \bibinfo {author} {\bibfnamefont {A.~A.}\ \bibnamefont {Kamel}},\ }\bibfield  {title} {\bibinfo {title} {Quasi-periodic orbits about the translunar libration point},\ }\href {https://doi.org/https://doi.org/10.1007/BF01227511} {\bibfield  {journal} {\bibinfo  {journal} {Celestial mechanics}\ }\textbf {\bibinfo {volume} {7}},\ \bibinfo {pages} {458} (\bibinfo {year} {1973})}\BibitemShut {NoStop}%
\bibitem [{\citenamefont {Kirillov}\ and\ \citenamefont {Levi}(2016)}]{kirillov2016_saddle}%
  \BibitemOpen
  \bibfield  {author} {\bibinfo {author} {\bibfnamefont {O.~N.}\ \bibnamefont {Kirillov}}\ and\ \bibinfo {author} {\bibfnamefont {M.}~\bibnamefont {Levi}},\ }\bibfield  {title} {\bibinfo {title} {Rotating saddle trap as foucault's pendulum},\ }\href {https://doi.org/https://doi.org/10.1119/1.4933206} {\bibfield  {journal} {\bibinfo  {journal} {American Journal of Physics}\ }\textbf {\bibinfo {volume} {84}},\ \bibinfo {pages} {26} (\bibinfo {year} {2016})}\BibitemShut {NoStop}%
\bibitem [{\citenamefont {Ucar}(2021)}]{ucar2021_polarity}%
  \BibitemOpen
  \bibfield  {author} {\bibinfo {author} {\bibfnamefont {H.}~\bibnamefont {Ucar}},\ }\bibfield  {title} {\bibinfo {title} {Polarity free magnetic repulsion and magnetic bound state},\ }\href {https://doi.org/https://doi.org/10.3390/sym13030442} {\bibfield  {journal} {\bibinfo  {journal} {Symmetry}\ }\textbf {\bibinfo {volume} {13}},\ \bibinfo {pages} {442} (\bibinfo {year} {2021})}\BibitemShut {NoStop}%
\bibitem [{\citenamefont {Committee}(2023)}]{IPT2023}%
  \BibitemOpen
  \bibfield  {author} {\bibinfo {author} {\bibfnamefont {I.}~\bibnamefont {Committee}},\ }\href@noop {} {\bibinfo {title} {Ipt 2023 problem list}},\ \bibinfo {howpublished} {\url{https://2023.iptnet.info/problems/}} (\bibinfo {year} {2023}),\ \bibinfo {note} {accessed: 27 May 2025}\BibitemShut {NoStop}%
\bibitem [{\citenamefont {Committee}(2024)}]{IYPT2024}%
  \BibitemOpen
  \bibfield  {author} {\bibinfo {author} {\bibfnamefont {I.}~\bibnamefont {Committee}},\ }\href@noop {} {\bibinfo {title} {Problems for the 37th iypt 2024}},\ \bibinfo {howpublished} {\url{https://www.iypt.org/problems/problems-iypt-2024/}} (\bibinfo {year} {2024}),\ \bibinfo {note} {accessed: 27 May 2025}\BibitemShut {NoStop}%
\bibitem [{\citenamefont {{Magnetic Games}}(2021)}]{youtube2021maglev}%
  \BibitemOpen
  \bibfield  {author} {\bibinfo {author} {\bibnamefont {{Magnetic Games}}},\ }\href {https://www.youtube.com/watch?v=V5FyFvgxUhE} {\bibinfo {title} {Maglev train with powerful magnets}} (\bibinfo {year} {2021}),\ \bibinfo {note} {accessed: 2025-06-27}\BibitemShut {NoStop}%
\bibitem [{\citenamefont {Hermansen}\ \emph {et~al.}(2023)\citenamefont {Hermansen}, \citenamefont {Durhuus}, \citenamefont {Frandsen}, \citenamefont {Beleggia}, \citenamefont {Bahl},\ and\ \citenamefont {Bj{\o}rk}}]{hermansen2023_rot}%
  \BibitemOpen
  \bibfield  {author} {\bibinfo {author} {\bibfnamefont {J.~M.}\ \bibnamefont {Hermansen}}, \bibinfo {author} {\bibfnamefont {F.~L.}\ \bibnamefont {Durhuus}}, \bibinfo {author} {\bibfnamefont {C.}~\bibnamefont {Frandsen}}, \bibinfo {author} {\bibfnamefont {M.}~\bibnamefont {Beleggia}}, \bibinfo {author} {\bibfnamefont {C.~R.}\ \bibnamefont {Bahl}},\ and\ \bibinfo {author} {\bibfnamefont {R.}~\bibnamefont {Bj{\o}rk}},\ }\bibfield  {title} {\bibinfo {title} {Magnetic levitation by rotation},\ }\href {https://doi.org/https://doi.org/10.1103/PhysRevApplied.20.044036} {\bibfield  {journal} {\bibinfo  {journal} {Physical Review Applied}\ }\textbf {\bibinfo {volume} {20}},\ \bibinfo {pages} {044036} (\bibinfo {year} {2023})}\BibitemShut {NoStop}%
\bibitem [{\citenamefont {Le~Lay}\ \emph {et~al.}(2024)\citenamefont {Le~Lay}, \citenamefont {Layani}, \citenamefont {Daerr}, \citenamefont {Berhanu}, \citenamefont {Dolbeault}, \citenamefont {Person}, \citenamefont {Roussille},\ and\ \citenamefont {Taberlet}}]{le2024_stable}%
  \BibitemOpen
  \bibfield  {author} {\bibinfo {author} {\bibfnamefont {G.}~\bibnamefont {Le~Lay}}, \bibinfo {author} {\bibfnamefont {S.}~\bibnamefont {Layani}}, \bibinfo {author} {\bibfnamefont {A.}~\bibnamefont {Daerr}}, \bibinfo {author} {\bibfnamefont {M.}~\bibnamefont {Berhanu}}, \bibinfo {author} {\bibfnamefont {R.}~\bibnamefont {Dolbeault}}, \bibinfo {author} {\bibfnamefont {T.}~\bibnamefont {Person}}, \bibinfo {author} {\bibfnamefont {H.}~\bibnamefont {Roussille}},\ and\ \bibinfo {author} {\bibfnamefont {N.}~\bibnamefont {Taberlet}},\ }\bibfield  {title} {\bibinfo {title} {Magnetic levitation in the field of a rotating dipole},\ }\href {https://doi.org/https://doi.org/10.1103/PhysRevE.110.045003} {\bibfield  {journal} {\bibinfo  {journal} {Physical Review E}\ }\textbf {\bibinfo {volume} {110}},\ \bibinfo {pages} {045003} (\bibinfo {year} {2024})}\BibitemShut {NoStop}%
\bibitem [{\citenamefont {Merkin}(2012)}]{merkin2012_introduction}%
  \BibitemOpen
  \bibfield  {author} {\bibinfo {author} {\bibfnamefont {D.~R.}\ \bibnamefont {Merkin}},\ }\href@noop {} {\emph {\bibinfo {title} {Introduction to the Theory of Stability}}},\ Vol.~\bibinfo {volume} {24}\ (\bibinfo  {publisher} {Springer Science \& Business Media},\ \bibinfo {year} {2012})\BibitemShut {NoStop}%
\bibitem [{\citenamefont {Bardin}\ and\ \citenamefont {Chekin}(2008)}]{bardin2008SatOrbital}%
  \BibitemOpen
  \bibfield  {author} {\bibinfo {author} {\bibfnamefont {B.}~\bibnamefont {Bardin}}\ and\ \bibinfo {author} {\bibfnamefont {A.}~\bibnamefont {Chekin}},\ }\bibfield  {title} {\bibinfo {title} {Orbital stability of planar oscillations of a satellite in a circular orbit},\ }\href {https://doi.org/https://doi.org/10.1134/S0010952508030118} {\bibfield  {journal} {\bibinfo  {journal} {Cosmic Research}\ }\textbf {\bibinfo {volume} {46}},\ \bibinfo {pages} {273} (\bibinfo {year} {2008})}\BibitemShut {NoStop}%
\bibitem [{\citenamefont {Wesson}(1978)}]{wesson1978plasmaStab}%
  \BibitemOpen
  \bibfield  {author} {\bibinfo {author} {\bibfnamefont {J.}~\bibnamefont {Wesson}},\ }\bibfield  {title} {\bibinfo {title} {Hydromagnetic stability of tokamaks},\ }\href {https://doi.org/https://doi.org/10.1088/0029-5515/18/1/010} {\bibfield  {journal} {\bibinfo  {journal} {Nuclear Fusion}\ }\textbf {\bibinfo {volume} {18}},\ \bibinfo {pages} {87} (\bibinfo {year} {1978})}\BibitemShut {NoStop}%
\bibitem [{\citenamefont {Balbus}\ and\ \citenamefont {Hawley}(1992)}]{balbus1992magDisk}%
  \BibitemOpen
  \bibfield  {author} {\bibinfo {author} {\bibfnamefont {S.~A.}\ \bibnamefont {Balbus}}\ and\ \bibinfo {author} {\bibfnamefont {J.~F.}\ \bibnamefont {Hawley}},\ }\bibfield  {title} {\bibinfo {title} {A powerful local shear instability in weakly magnetized disks. iv. nonaxisymmetric perturbations},\ }\href {https://doi.org/https://ui.adsabs.harvard.edu/link_gateway/1992ApJ...400..610B/doi:10.1086/172022} {\bibfield  {journal} {\bibinfo  {journal} {Astrophysical Journal v. 400, p. 610-621}\ }\textbf {\bibinfo {volume} {400}},\ \bibinfo {pages} {610} (\bibinfo {year} {1992})}\BibitemShut {NoStop}%
\bibitem [{\citenamefont {Meerkov}(1980)}]{meerkov1980VibControl}%
  \BibitemOpen
  \bibfield  {author} {\bibinfo {author} {\bibfnamefont {S.}~\bibnamefont {Meerkov}},\ }\bibfield  {title} {\bibinfo {title} {Principle of vibrational control: theory and applications},\ }\href {https://doi.org/https://doi.org/10.1109/TAC.1980.1102426} {\bibfield  {journal} {\bibinfo  {journal} {IEEE Transactions on Automatic Control}\ }\textbf {\bibinfo {volume} {25}},\ \bibinfo {pages} {755} (\bibinfo {year} {1980})}\BibitemShut {NoStop}%
\bibitem [{\citenamefont {Murray}\ \emph {et~al.}(2017)\citenamefont {Murray}, \citenamefont {Li},\ and\ \citenamefont {Sastry}}]{murray2017robControl}%
  \BibitemOpen
  \bibfield  {author} {\bibinfo {author} {\bibfnamefont {R.~M.}\ \bibnamefont {Murray}}, \bibinfo {author} {\bibfnamefont {Z.}~\bibnamefont {Li}},\ and\ \bibinfo {author} {\bibfnamefont {S.~S.}\ \bibnamefont {Sastry}},\ }\href {https://doi.org/https://doi.org/10.1201/9781315136370} {\emph {\bibinfo {title} {A mathematical introduction to robotic manipulation}}}\ (\bibinfo  {publisher} {CRC press},\ \bibinfo {year} {2017})\BibitemShut {NoStop}%
\end{thebibliography}%

\end{document}